\documentclass[11pt]{article}
\usepackage{jcappub,natbib,float,caption}

\usepackage{graphicx,amsmath,amssymb,color,bm}
\usepackage[caption=false]{subfig}

\def\be{\begin{equation}}
\def\ee{\end{equation}}
\def\bea{\begin{eqnarray}}
\def\eea{\end{eqnarray}}
\newcommand{\vs}{\nonumber\\}
\def\ba#1\ea{\begin{align}#1\end{align}}
\def\bg#1\eg{\begin{gather}#1\end{gather}}

\newcommand{\s}{\sigma}
\newcommand{\refeq}[1]{Eq.~(\ref{eq:#1})}          
\newcommand{\refeqs}[2]{Eqs.~(\ref{eq:#1})--(\ref{eq:#2})}

\newcommand{\refsec}[1]{Sec.~\ref{sec:#1}}          
\newcommand{\refapp}[1]{App.~\ref{app:#1}}

\renewcommand{\v}[1]{\mathbf{#1}}

\newcommand{\vx}{\v{x}}
\newcommand{\vv}{\v{v}}

\newcommand{\vk}{\v{k}}
\newcommand{\vq}{\v{q}}

\renewcommand{\d}{\delta}

\newcommand{\eps}{\epsilon}

\newcommand{\vn}{\v{\nabla}}
\newcommand{\Om}{\Omega_m}

\def\O{\mathcal{O}}

\def\Del{\mathcal{D}}

\newcommand{\Tr}{{\rm Tr}}

\def\cH{\mathcal{H}}

\newcommand{\comment}[1]{}

\def\x{{\bf x}}
\def\q{{\bf q}}

\def\xfl{\v{x}_{\rm fl}}

\newcommand\dbi[1]{\delta_{\rm lgr}^{[#1]}(\vq)}

\def\lgr{{\rm lgr}}

\def\der{\partial}
\def\vf{\v{f}}
\def\vep{\varepsilon}
\def\s{\v{s}}

\def\Dc{{\rm D}}
\def\dc{{\rm d}}
\def\vp{\v{p}}

\title{Biased Tracers and Time Evolution}

\author[a]{Mehrdad Mirbabayi,}

\author[b]{Fabian Schmidt,}

\author[a]{Matias Zaldarriaga}

\affiliation[a]{Institute for Advanced Study, Princeton, NJ~08540, USA}

\affiliation[b]{Max-Planck-Institut f\"ur Astrophysik, Karl-Schwarzschild-Str. 1, 85741 Garching, Germany}

\abstract{
We study the effect of time evolution on galaxy bias. We argue that at any order in perturbations, the galaxy density contrast can be expressed in terms of a finite set of locally measurable operators made of spatial and temporal derivatives of the Newtonian potential. This is checked in an explicit third order calculation. There is a systematic way to derive a basis for these operators.  This basis spans a larger space than the expansion in gravitational and velocity potentials usually employed, although new operators only appear at fourth order.   The basis is argued to be closed under renormalization. Most of the arguments also apply to the structure of the counter-terms in the effective theory of large-scale structure. 
}

\begin{document}

\maketitle
\flushbottom

\section{Introduction}

The purpose of the theory of bias is to relate observable properties of tracers such as the density contrast of galaxies $\delta_g$ to the underlying matter distribution and the initial conditions.  If one neglects primordial non-Gaussianity and possible entropic perturbations, as we will do throughout for simplicity, then $\delta_g$ can only be a function of local observables of the dark-matter field (such as density, tidal forces, etc.) along the tracer's trajectory. In the same approximation the deviation of the tracer's trajectory from that of the underlying cosmic fluid, typically phrased as \emph{velocity bias}, should also depend only on those locally observable quantities.

A standard approach in the field has been to try to write the most general functional dependence of this sort at the final time \cite{mcdonald/roy:2009,Assassi} ({\em Eulerian biasing}). The ignorance about the underlying physics is then parameterized into a set of bias coefficients which become less and less relevant (in a perturbation theory sense) as we go to higher and higher orders in matter fluctuations and derivatives. An alternative approach is to perform the parametrization on the initial slice \cite{mo/white:1996,matsubara:2008,PBSpaper} (\emph{Lagrangian biasing}). However, the sufficiency of either of these approaches is not totally obvious: It is true that because of deterministic evolution the full formation history can be reconstructed from the information on either the final or the initial time slice. However, because of the long-range gravitational interactions, one may suspect that the Eulerian (Lagrangian) description of galaxy distribution for a population of finite age might be non-local. Regardless of this issue, a complete and non-redundant classification of locally observable operators does not seem to exist so far. 

Finally in \cite{Senatore:2014eva} it was argued that the most general relation between tracers and the underlying matter distribution should be written as an integral over time along the fluid trajectory of all possible local observables including the tidal tensor, gradients of the velocity field and other operators with higher spatial derivatives. The relation is thus local in space but non-local in time. In \cite{Senatore:2014eva} it was shown that the time-non-locality can be dealt with in perturbation theory. 

The goal of this paper is to investigate systematically the implications of the finite age and time evolution of tracers on the effective bias parametrization. In particular we will show that all the approaches described above are mathematically equivalent and provide a dictionary between them. We will give a complete and yet non-redundant basis of operators that can be used to express the most general relation between biased tracers and the underlying matter fluctuations.  We are not the first who study this problem. Considerable insight has been obtained starting from the work of \citet{Tegmark}. In particular generation of nonlinear bias due to time evolution was first discussed in \cite{chan/scoccimarro/sheth:2012}. Evidence from numerical simulations as well as phenomenological explanations have been provided in \cite{Baldauf,Sheth}.

\section{A Green's function for galaxy formation}
\label{sec:basics}

In order to study this problem, we consider instantaneous formation of tracers at time $\tau_*$ and write the most general expression for $\delta_g(\tau_*)$ in terms of local observables, $\{O(\tau_*)\}$, made of the underlying matter distribution with coefficients $\{b_{O}^*\}$:
\be
\d_g(\vx,\tau_*) = \sum_{O} b_{O}^* O(\vx, \tau_*)\,.
\label{eq:dgstar}
\ee
Postulating the conservation of the tracers we then solve for $\delta_g(\tau)$ at later times. We shall see that this provides a Green's function for galaxy biasing. Any formation history which depends on locally measurable quantities can be obtained by a superposition of these conserved tracers. A similar approach was used in \cite{Senatore:2014eva}.

One should ask what are the locally measurable quantities that instantaneously forming objects can depend upon.  As long as the tracers are well described as a non-relativistic fluid,  the gravitational effects are fully captured by the potential $\phi$ (i.e. the perturbation to the time-time component of the metric), which for convenience is rescaled to $\Phi\equiv 2\phi/3\cH^2\Omega_m$, satisfying $\nabla^2 \Phi=\delta$. Because of the equivalence principle, only tidal forces made of various powers of $\der_i\der_j\Phi$ are measurable; a uniform, but possibly time-dependent, gravitational force $\partial_i \Phi(t)$ causes a uniform acceleration of all objects and hence cannot be seen by any local measurement. It follows that locally measurable quantities have to be invariant under time-dependent but uniform translations.

In addition to $\der_i\der_j\Phi$, its spatial derivatives as well as time derivatives along the flow of dark matter are also locally measurable. Throughout, we always work with this \emph{convective time derivative} defined through
\be
\frac{\rm D}{\rm d\tau} O(\vx,\tau) \equiv 
\left[\der_\tau + \vv\cdot\der_x\right] O(\vx,\tau)\,,
\ee
where $\vv(\vx,\tau)$ is the matter peculiar velocity field.\footnote{In appendix \refapp{dv} we will show that velocity dispersion $\der_i v_j$, which is also locally measurable, can be expressed in terms of $\der_i\der_j\Phi$ and its time-derivatives.}  

Since $\der_i\der_j\Phi$ is dimensionless, additional derivatives must be accompanied by the characteristic size and formation time scale of the object which are denoted respectively by $L$ and $T$.  The scale $L$ is naively expected to be that which encloses the mass of the object at the background density, i.e. the Lagrangian radius of the object, although baryonic feedback and radiative transfer effects could in principle increase this scale for certain tracers. The time scale $T$ on the other hand is expected to be typically the Hubble time $T \sim \cH^{-1}(\tau_*)$, the same scale that is used to define $\Phi$.  
Higher order spatial derivatives associated with the length
scale $L$ are not directly relevant for the issue of time evolution which
we are concerned with here.  Thus, throughout the paper we will only consider
terms that are lowest order in spatial derivatives.  This means that each instance
of $\Phi$ will come with exactly two spatial derivatives.

We are therefore interested in the evolution of local observables along the large-scale flow of dark-matter. (We assume zero velocity bias for the moment. It is considered later and shown to be higher order in spatial derivatives.) This problem is naturally posed in the Lagrangian formalism where fluid trajectories are labeled by their initial position $\vq$ and are implicitly defined by
\be
\vx_{\rm fl}(\vq,\tau) = \vq + \v{s}(\vq,\tau)\,,
\ee
where $\v{s}(\vq,0) = 0$.  The displacement then satisfies
\be
\v{s}(\vq,\tau_0) = \int_0^{\tau_0} \vv(\vx_{\rm fl}(\vq, \tau), \tau) \dc\tau\,.
\label{eq:sdef}
\ee
We will frequently drop the $\vq$ argument in $\vx_{\rm fl}$ when there is no ambiguity.

The system to solve for $\tau > \tau_*$ consists of the continuity equations for tracer and
matter in addition to the Euler equation,
\bea
\label{eq:cont2}
\frac{\Dc}{\dc\tau} \d  
&=\:& - \theta (1+\delta),  \\
\frac{\Dc}{\dc\tau} \theta& =\:& - \cH \theta - \frac32 \Om \cH^2 \d  - (\partial_x^i \vv^j)^2\,,
\label{eq:convSPT}\\
\frac{\Dc}{\dc\tau} \d_g &=\:& -\theta (1+\delta_g)
\label{eq:cont1}
\eea
where $\Dc/\dc\tau$ is the convective derivative and $\theta = \partial_x^i v_i$ is the dark matter velocity divergence.\footnote{We will drop the subscripts on $\der_x$ and $\der_q$ when the distinction is clear from the argument of the functions.}

\subsection{Flow of dark matter}

The first two equations are independent of the last one and describe the time evolution of dark matter. In the standard perturbation theory $\delta(\vx,\tau)$ and $\theta(\vx,\tau)$ are solved in powers of the initial field $\delta_1(\vx)$ evaluated at the same position $\vx$. For instance, $\delta^{[1]}(\vx,\tau) = D(\tau)\delta_1(\vx)$ and 
\be \label{eq:d2}
\delta^{[2]}(\vx,\tau)= D^2(\tau) \left[\frac{5}{7}\delta_1^2(\vx)
+\frac{2}{7} \left(\frac{\der_i\der_j}{\nabla^2}\delta_1(\vx)\right)^2
+\left(\frac{\der_i}{\nabla^2}\delta_1(\vx)\right)\der_i\delta_1(\vx)\right],
\ee
where $D(\tau)$ is the linear growth factor normalized to $a(\tau)$ during matter domination. We have made the usual very accurate approximation of replacing the $n$-th order growth factor with $D^n(\tau)$.

The last term in the above expression for $\delta^{[2]}$ blows up in the presence of an infinitely long-wavelength mode. The leading effect of the long wavelength mode is a large uniform acceleration, and the last term of \refeq{d2} is the displacement needed to move the argument of the linear solution from $\vx$ to the initial Lagrangian position: $\delta_1(\vx)=\delta_1(\vq)-D(\tau)(\der_i/\nabla^2)\delta_1(\vx)\der_i\delta_1(\vx)+\O(\delta_1^3)$. Similarly, the third order solution $\delta^{[3]}(\vx,\tau)$ is shown in appendix \refapp{delta3} to contain terms which diverge in the presence of infinite wavelength modes. They are needed to displace lower order solutions from $\vx$ to $\vq$. Therefore, expanding $\delta(\vx,\tau)$ in terms of the initial field at the Lagrangian position $\delta_1(\vq)$ yields 
\be
\label{eq:delta}
\delta(\vx,\tau)= D(\tau) \dbi1 + D^2(\tau) \dbi2 +D^3(\tau)\dbi3+\cdots
\ee
where now $\dbi2= 5/7 \delta_1^2(\vq)+2/7 [(\der_i\der_j/\nabla^2)\delta_1(\vq)]^2$ and $\dbi3$, given below, are {\em Lagrangian} operators constructed from the initial density field $\dbi1=\delta_1(\vq)$. They remain finite in the presence of long wavelength modes. In fact, the structure of \refeq{cont2} and \refeq{convSPT} ensures that this holds to all orders, since (a) after expressing $\delta$ and $\theta$ as a function of $(\vq,\tau)$ the convective derivative reduces to an ordinary time derivative, and then (b) every term in the equation can be written in terms of $\der_i\der_j\Phi$ or $\der_iv_j$, which remain finite for infinitely long-wavelength perturbations. In \refapp{convSPT} and independently in \refsec{Lagr} we provide methods of solving directly for $\dbi n$.\footnote{The implication of the Lagrangian expansion \eqref{eq:delta} on the form of the effective stress-tensor in the EFT of large scale structures was discussed in \cite{3pf}.}

What we mean by an $n^{th}$ order Lagrangian operator is that in momentum space there is no singularity when any subset of $n$ momenta $\{\vk_a\}$ is taken to zero, with $\{\delta_1(\vk_a)\}$ kept fixed. This doesn't ensure that the operator has a spatially local expression [even in terms of $\Phi_1(\vq)\equiv \nabla^{-2}\delta_1(\vq)$], as can be seen from the explicit solution
\be
\dbi3 =\: \frac{341}{567} \d_1^3 + \frac{11}{21} \d_1 K_1^2 + \frac 29 K_1^3 
+ \frac4{63} K_1^{ij} \Del_{ij} \left[  \d_1^2 - \frac32 K_1^2  \right]\,,
\label{eq:d3inv1}
\ee
where the derivative operator $\Del_{ij}$ and the traceless tidal tensor are defined respectively as
\be\label{eq:Del}
\Del_{ij} \equiv \left(\frac{\partial_i\partial_j}{\nabla^2} - \frac13 \d_{ij} \right); \quad
K^1_{ij}(\vq) \equiv \Del_{ij} \delta_1(\vq)\,.
\ee
The term in \refeq{d3inv1} involving $\Del_{ij} (\delta_1^2 - 3 K_1^2/2)$ cannot be written as a local combination of $\partial_i\partial_j\Phi_1$. Since this is an important distinction for the discussion in this paper, let us emphasize again: \emph{locally measurable observables} contain \emph{spatially nonlocal Lagrangian operators} in their expansion.

Beyond the density $\d(\vx,\tau)$, the equivalence principle ensures that any locally measurable quantity, including the tidal tensor $\der_i\der_j\Phi(\vx,\tau)$, can be solved along the flow in terms of such Lagrangian operators. Hence for any scalar $O$ involving $d_O$ powers of $\der_i\der_j\Phi$ we have:
\be
\label{eq:Obi}
O(\vx,\tau)=D^{d_O}(\tau) O_\lgr^{[d_O]}(\vq)+D^{d_O+1}(\tau) O_\lgr^{[d_O+1]}(\vq)+\cdots
\ee
where the first component $O_\lgr^{[d_O]}(\vq)$ is the same as $O(\vx,\tau)$ with the replacement 
\be
\der_i\der_j\Phi(\vx,\tau)\to (\der_i\der_j\Phi)_\lgr^{[1]}(\vq)\equiv \frac{\der_i\der_j}{\nabla^2}\delta_1(\vq).
\ee
Higher order $O_\lgr^{[n]}(\vq)$ are constructed from products of $d_O$ factors of $(\der_i\der_j\Phi)_\lgr^{[m_a]}(\vq)$ with $\sum_{a=1}^{d_O} m_a =n$. 

\subsection{Flow of galaxies}\label{sec:flg}

Let us now turn to \refeq{cont1} to solve for $\delta_g(\tau)$. It is useful to subtract \refeq{cont2} to get
\be
\frac{\Dc}{\dc\tau}\ (\delta_g -\delta)=-\theta (\delta_g-\delta).
\label{eq:D/dtdiff}
\ee
This is a linear and homogeneous equation in $\delta_g-\delta$, hence we can make use of the superposition principle. The solution $\delta_g(\tau)$ for an arbitrary formation history can be obtained by breaking the formation into instantaneous events and superposing the corresponding solutions.

We now show that if $\delta_g(\tau_*)$ has a bias expansion fully in terms of local operators as in \refeq{dgstar}, then at any later time it remains local:
\be
\d_g(\vx,\tau) = \sum_{O} b_{O}(\tau) O(\vx, \tau)\,,
\label{eq:dg}
\ee
where the sum is over all local operators, and $\{b_O(\tau)\}$ are uniquely determined in terms of $\{b_O^*\}$. We proceed now with the general derivation; an explicit calculation showing how this works in practice up to third order will appear below in \refsec{IF}.

Substituting \refeq{dg} as an ansatz for $\delta_g(\tau)$ and \refeq{Obi} for operators in \refeq{D/dtdiff}, and using the fact that convective derivative becomes an ordinary derivative when acted on functions of $(\vq,\tau)$, we obtain
\be
\label{eq:ode}
\sum_O\sum_{n=d_O}^\infty \left[\frac{\dc\tilde b_O}{\dc\tau}
+\tilde b_O\left(n\frac{\dc\log D(\tau)}{\dc\tau}+\sum_{m=1}^\infty D^m(\tau)\theta_\lgr^{[m]}(\vq)\right)\right]
D^n(\tau)O_\lgr^{[n]}(\vq) =0,
\ee
where $\tilde b_O \equiv b_O$ except for $\tilde b_\delta \equiv b_1 -1$. There is a lot of overlap among $O_\lgr^{[n]}(\vq)$ and their products for various $O(\vx,\tau)$ operators. Expressing everything in \refeq{ode} in terms of a linearly independent basis for Lagrangian operators $\{E_\lgr(\vq)\}$, and setting the coefficient of each one to zero yields a set of coupled first order ODE's for the bias parameters $\{b_{O}(\tau)\}$, with the initial condition $b_{O}(\tau_*)=b_O^*$.

Let us also expand local operators $\{O(\vx,\tau)\}$ in terms of a linearly independent Eulerian basis $\{\hat E(\vx,\tau)\}$, and introduce a bias parameter $b_{\hat E}(\tau)$ for each. The system is guaranteed to have a solution if every $E_\lgr(\vq)$ can be mapped to a unique local operator $\hat E(\vx,\tau)$ such that $\dc \tilde b_{\hat E}/\dc\tau$ multiplies $E_\lgr(\vq)$ in \refeq{ode}. At first sight this may seem not to be possible. Naively, there seems to be many $O_\lgr^{[n]}(\vq)$ for every $O(\vx,\tau)$ and as we saw in \refeq{d3inv1} they often have spatially non-local structures. This structure will appear in one of the $E_\lgr(\vq)$ and if there is no analogous locally measurable operator, the ansatz \eqref{eq:dg} would fail.

However, the non-local structures that appear in $O^{[n]}_\lgr(\vq)$ are not arbitrary. And indeed we can show that every $O_\lgr^{[n]}(\vq)$ can be mapped to a local Eulerian operator $\hat O^{[n]}(\vx,\tau)$ which is unique up to the addition of higher order local operators. Recall that the convective derivative of a locally measurable quantity is also locally measurable. Since in the expansion \refeq{Obi} every $O_\lgr^{[n]}$ is accompanied by a different power of $D(\tau)$, by combining convective time derivatives of $O(\vx,\tau)$ we can easily construct the desired operator 
\be
\hat O^{[n]}(\vx,\tau)=D^n(\tau) O_\lgr^{[n]}(\vq)+\O(\delta^{n+1}).
\ee
For instance, $\hat O^{[d_O]}(\vx,\tau) \equiv  O(\vx,\tau)$, and at next order:
\be
\hat O^{[d_O+1]}(\vx,\tau) \equiv  O'(\vx,\tau)-d_O O(\vx,\tau),
\ee
where prime denotes convective derivative with respect to $\log D(\tau)$:
\be
O'(\vx,\tau) = \frac{\Dc}{\dc\log D(\tau)} O(\vx,\tau)=\frac{D(\tau)}{\dot D(\tau)}\frac{\Dc}{\dc\tau} O(\vx,\tau).
\ee
At higher orders $\hat O^{[n]}(\vx,\tau)$ can be constructed recursively:
\be
\hat O^{[n]} \equiv \frac{1}{(n-d_O)!} \left[ (\hat O^{[n-1]})'- (n-1)\hat O^{[n-1]}\right].
\ee
The above mapping leads to a one to one mapping between the linearly independent bases of operators $\{E_\lgr(\vq)\}$ and $\{\hat E(\vx,\tau)\}$, and confirms the sufficiency of the ansatz \refeq{dg}.

To lowest order in spatial derivatives, a complete and non-redundant basis $\{E_\lgr(\vq)\}$ for all $\{O_\lgr^{[n]}\}$, is made of all scalars that can be constructed from products of $\Pi^{[n]}_{ij}(\vq)\equiv (\der_i\der_j\Phi)_\lgr^{[n]}(\vq)$. By the equations of motion $\Tr [\Pi_{ij}^{[n]}]$ can be expressed in terms of lower order traces and hence is not independent. We therefore have up to fourth order (and using matrix notation)
\bea
{\rm 1^{st}} \ && \ {\rm Tr}[\Pi^{[1]}]  \label{eq:listP} \\ 
{\rm 2^{nd}} \ && \ {\rm Tr}[(\Pi^{[1]})^2],\  ({\rm Tr}[\Pi^{[1]}])^2 \nonumber\\ 
{\rm 3^{rd}} \ && \ {\rm Tr}[(\Pi^{[1]})^3 ],\ {\rm Tr}[(\Pi^{[1]})^2 ]  {\rm Tr}[\Pi^{[1]}],\ ({\rm Tr}[\Pi^{[1]}])^3,\ {\rm Tr}[\Pi^{[1]} \Pi^{[2]}] \nonumber \\ 
{\rm 4^{th}} \ && \ {\rm Tr}[(\Pi^{[1]})^4 ],\ {\rm Tr}[(\Pi^{[1]})^3 ]  {\rm Tr}[\Pi^{[1]}],\  {\rm Tr}[(\Pi^{[1]})^2 ]  {\rm Tr}[(\Pi^{[1]})^2],\ ({\rm Tr}[\Pi^{[1]}])^4,\ {\rm Tr}[\Pi^{[1]} \Pi^{[3]}], \nonumber\\
&& \   {\rm Tr}[\Pi^{[2]} \Pi^{[2]}]. \nonumber
\eea
In practice it is more natural to work with the distortion matrix $M^{ij} =\der_q^i s^j$ in the Lagrangian space. In \refsec{Lagr} we will discuss the connection between the two sets of variables and write an identically looking basis in terms of various $M_{ij}^{[n]}(\vq)$. 

To obtain the corresponding Eulerian basis $\{\hat E(\vx,\tau)\}$, we define the analogs of $\Pi_{ij}^{[n]}(\vq)$. The time-derivatives of $\der_i\der_j\Phi(\vx,\tau)$ can be combined to obtain Eulerian operators $\hat\Pi_{ij}^{[n]}(\vx,\tau)$ whose leading order expression in perturbation theory are given by $\hat\Pi_{ij}^{[n]}(\vx,\tau) = D^n(\tau) \Pi_{ij}^{[n]}(\vq)+\O(\delta^{n+1})$. Starting from
\be
\hat\Pi^{[1]}_{ij}(\vx,\tau) =\: \der_i\der_j\Phi(\vx,\tau),
\ee
the higher order operators can again be calculated recursively. Suppressing the indices and the argument $(\vx,\tau)$
\be
\hat\Pi^{[n]} =\:\frac{1}{(n-1)!} \left[(\hat\Pi^{[n-1]})' - (n-1)\hat\Pi^{[n-1]}\right].
\ee
The Eulerian basis is then composed of all scalars constructed out of $\hat\Pi^{[n]}_{ij}$ excluding ${\rm Tr}[\hat\Pi^{[n]}]$; in other words, it can be obtained by replacing $\Pi$ with $\hat\Pi$ in \refeq{listP}. We conclude by a few comments:

\begin{itemize}

\item The tensors $\hat \Pi^{[n]}(\vx,\tau)$, and hence the Eulerian operators $\{\hat E(\vx,\tau)\}$, are easily calculable in momentum space by combining the $F$ and $G$ kernels of the Standard Perturbation Theory (SPT). For instance,
\be
\hat\Pi^{[1]}_{ij}(\vk,\tau) =\:\frac{\vk^i \vk^j}{|\vk|^2}\sum_{n=1}^{\infty}D^n(\tau) \delta_n(\vk),
\ee
where $\delta_n(\vk)$ is the $n^{th}$ order SPT solution, and 
\be
\begin{split}
\hat\Pi^{[2]}_{ij}(\vk,\tau) =\:\sum_{n=1}^{\infty}D^n(\tau)
\Big[&\frac{\vk^i \vk^j}{|\vk|^2} (n-1)\delta_n(\vk)\\[8pt]
&+\sum_{m=1}^{n-1}\int\frac{\dc^3\vp}{(2\pi)^3}\frac{(\vk-\vp)\cdot \vp}{|\vk-\vp|^2}\frac{\vp^i \vp^j}{|\vp|^2}\theta_m(\vk-\vp)\delta_{n-m}(\vp)\Big],
\end{split}
\ee
and so on.

\item One should bear in mind that although $\der_i\der_j\Phi(\vx,\tau)$ is a total derivative, neither its convective time derivatives nor the $\vq$-space matrices $\Pi_{ij}^{[n]}(\vq)$ have necessarily to be full spatial derivatives of any kind.

\item The list \eqref{eq:listP} spans only the subset of Lagrangian operators that appear in the expansion of local operators. There are infinitely many more Lagrangian operators. For instance, at third order one may expect at least five operators:
\be
\delta_1^3,\quad \delta_1K_1^2,\quad K_1^3,\quad K_1^{ij} \Del_{ij} \d_1^2,\quad K_1^{ij} \Del_{ij} K_1^2,
\ee
but there are four in \refeq{listP}. The last two operators always appear with relative factor $-3/2$ in the expansion of local operators, as in \refeq{d3inv1}. 

\item If $\{b_{O}^*\}$ were known we could obtain $\{b_{O}(\tau)\}$ by solving \refeq{ode}.\footnote{The solutions is unique only if we use a linearly independent basis of operators like $\{\hat E(\vx,\tau)\}$. In the following this is always implied.} Even if $b_O(\tau_*)=0$ for some operator, generically it will be non-zero after time evolution. (Examples of this will be seen in the next section.) In the absence of that information, an effective Eulerian bias can be fully formulated in terms of local operators with unknown coefficients.

\item Up to third order our Eulerian list consists of seven operators and agrees with \cite{mcdonald/roy:2009,chan/scoccimarro/sheth:2012,Assassi}.\footnote{There is one redundant third order operator in \cite{mcdonald/roy:2009}.} Starting from third order convective time derivatives are needed (i.e. our $\hat\Pi^{[2]}(\vx,\tau)$ operator.) In terms of the commonly used velocity potential $\nabla^2 \Phi_v=-\cH^{-1}\theta$, we have
\be
\hat\Pi^{[2]}_{ij} = \der_i\der_m\Phi\der_j\der_m\Phi+\frac{5}{2}(\der_i\der_j\Phi-\der_i\der_j\Phi_v)+\O(\delta^3).
\ee
Clearly, at higher orders $\der_i\der_j\Phi$ and $\der_i\der_j\Phi_v$ will be insufficient for a local description of bias and higher order time derivatives will be necessary. The first example is the quartic operator $\Tr(\hat\Pi^{[1]} \hat\Pi^{[3]})$. This contributes to five-point correlation function of $\delta_g$ at tree level, and the power-spectrum at two loops.

\end{itemize}

\subsection{Bias evolution at third order: explicit calculation}
\label{sec:IF}

In this section, we explicitly derive the density field of tracers $\d_g$ at time $\tau$ given formation at some earlier time $\tau_*$, up to
third order in perturbation theory.  
This will illustrate the general time-dependence of the bias parameters, and confirms the arguments of last section. 
For this calculation, we use an approach we call ``convective SPT'', which is a hybrid
of the  standard Eulerian perturbation theory (SPT) and Lagrangian
perturbation theory (LPT) approaches.  The details are provided in \refapp{convSPT}.\footnote{\citet{chan/scoccimarro/sheth:2012} have presented a similar calculation, and we reach agreement with their results.  Note however that what is referred to as nonlocal terms in \cite{chan/scoccimarro/sheth:2012}, e.g. $(K_{ij})^2$, is part of our local observables. Third order bias has been recently compared with $N$-body simulations in \cite{Saito,Biagetti}.} 

At $\tau_*$, we start from the most general, local, third order expression for $\delta_g(\tau_*)$ \cite{chan/scoccimarro/sheth:2012}: 
\be
\begin{split}
\d_g^* = \d_g(\xfl(\tau_*),\tau_*) =\:  \sum_{n=1}^3 &\frac{b_n^*}{n!} [\d^*]^n 
+ \sum_{n=2}^3 \frac{b_{K^n}^*}{n!} \Tr \left[ (K_{ij}^*)^n \right]\\[5pt]
&+ \frac16 b^*_{\d K^2} \d^*\: \Tr \left[ (K_{ij}^*)^2 \right] +\frac{1}{6} b^*_{\Gamma} \Gamma^* 
+ \eps^*\,,
\end{split}
\label{eq:dgIC}
\ee
where all quantities are measured at $(\xfl(\tau_*), \tau_*)$. We have defined
\ba
K_{ij}(\vx,\tau) \equiv\: &\hat\Pi^{[1]}_{ij}-\frac{1}{3}\delta_{ij}\Tr (\hat\Pi^{[1]}),\\[10pt]
\Gamma(\vx,\tau) \equiv\: &\frac{21}{10} K^{ij}(\hat\Pi^{[2]}_{ij}-\hat\Pi^{[1]}_{ik}\hat\Pi^{[1]}_{jk}).
\ea
Using the definition \eqref{eq:Del}, the leading Lagrangian component of $\Gamma$ is
\be
\label{eq:nl}
\Gamma^{[3]}_\lgr(\vq) = K_1^{ij} \Del_{ij} \left[  \d_1^2 - \frac32 K_1^2  \right]
\ee
In the second line of \refeq{dgIC}, we have introduced
a stochastic variable $\eps^*(\vx)$ (considered first order in perturbations)
which is defined on the formation
time slice and captures the fluctuations in galaxy density that are
uncorrelated with long-wavelength perturbations.  However, the statistical properties 
of the stochastic term in general also depend on local observables. Thus, in order to keep track of the effect of stochasticity up to cubic order in fluctuations we write:
\be
\eps^* = \eps^*_0 + \eps^*_{\d} \d^* + \frac12 \eps^*_{\d^2} [\d^*]^2
+ \frac12 \eps^*_{K^2} \Tr \left[ (K_{ij}^*)^2 \right] + \cdots
\ee
where $\eps^*_\alpha$ are stochastic variables. At the order we are working in, for two-point correlation functions one needs to know $\langle \eps^*_0 \eps^*_0 \rangle$ and three other correlations to describe the dependence on the long wavelength modes:  $\langle\eps^*_{\d}\eps^*_{\d} \rangle$, $\langle \eps^*_0 \eps^*_{\d^2} \rangle$ and $\langle \eps^*_0 \eps^*_{K^2} \rangle$. 

We solve \refeq{D/dtdiff} as described in \refapp{convSPT}, and arrive at
the following expression at later times along the fluid trajectory:
\ba
\d_{g}(\xfl(\tau),\tau) =\:&\sum_{n=1}^3 \frac{b_n^E}{n!} \d^n 
+ \sum_{n=2}^3 \frac{b_{K^n}^E}{n!} \Tr \left[ K_{ij}^n \right]
+ \frac16 b_{\d K^2}^E \d\: \Tr \left[ K_{ij}^2 \right] + \frac16 b_{\Gamma}^E \Gamma \label{eq:dgE}\\
& +\eps^*_0 + \eps_\d^E \d + \frac12 \eps_{\d^2}^E \d^2 + \frac12 \eps_{K^2}^E \Tr \left[ K_{ij}^2 \right] \,,\nonumber
\ea
where the quantities on the r.h.s. are evaluated at $(\xfl(\tau), \tau)$. Note that as anticipated the list of local operators is sufficient to express the time-evolved $\delta_g(\tau)$. 

The Eulerian bias coefficients $b_O^E$ are given in terms of $b^*_O$ in \refeq{bE2} and \refeq{biasEthirdorder}. Let us quote some of them for reference:
\ba
\label{eq:b_1E}
b_1^E(\tau) =\:& (b_1^* - 1)D_* + 1, \\
b_2^E(\tau) =\:& b_2^* D_*^2 - \frac8{21} (b_1^*-1) D_* (D_*-1),\\
\label{eq:bK2}
b_{K^2}^E(\tau) =\:& b_{K^2}^* D_*^2 + \frac47 (b_1^*-1) D_* (D_*-1),\\
\label{eq:bEnl}
b_{\Gamma}^E(\tau) =\:& b_\Gamma^* D_*^3 +\left[
(b_1^*-1) \left(-\frac8{21}\right) \left(\frac{23}7 - D_*\right) 
+\frac{20}7 b_{K^2}^* D_* 
\right]
D_* (D_*-1)
\ea
where $D_*= D(\tau_*)/D(\tau)$. The Eulerian stochasticity terms $\eps^E_\alpha$ are given in \refeq{epsE2} and \refeq{epsEthirdorder}.

In general, obtaining these bias coefficients requires a detailed calculation. However there is a subset of operators for which obtaining the time dependence of the bias parameters is rather straightforward. The key observation is that the r.h.s. of \refeq{D/dtdiff} is a product two scalar operators $\theta(\delta_g-\delta)$. If we substitute the bias expansion \eqref{eq:dg} and expand in terms of a linearly independent Lagrangian basis of operators, every operator on the r.h.s. would be {\em factorizable}, i.e. it can be written as a product of two scalars. Thus, \emph{any term in $\delta_g -\delta$ that cannot be written as a product of two scalar operators must be time-independent.} 

Suppose $O(\vx,\tau)$ is non-factorizable (for instance it can be $\delta$ or $K_{ij}^2$), then so is $O_\lgr^{[d_O]}(\vq)$. If $O_\lgr^{[d_O]}(\vq)$ does not appear in the Lagrangian solution for any other operator, for its coefficient to be time-independent, we must have 
\be
\label{eq:bsol}
b_O(\tau) = b_O^* D_*^{d_O}(\tau)  \qquad \text{with}\qquad 
D_*(\tau) = \frac{D(\tau_*)}{D(\tau)}\,.
\ee
For instance consider the case where $O$ is just $\delta$. Then we conclude that $(b_1-1)D(\tau)\delta_1(\vq)$ is time independent because the r.h.s. of \refeq{D/dtdiff} starts at second order. As a result, $(b_1-1)\propto {D^{-1}(\tau)}$, as implied by \refeq{bsol}. This agrees with \refeq{b_1E}. If $O^{[d_O]}_\lgr$ does appear in the expansion of some lower order operator $P$, then by superposition principle we can still write \refeq{bsol} plus corrections proportional to $b_{P}^*$. For instance, If $O = K^2$ then $(K^2)^{[2]}_\lgr=K_1^2(\vq)$ also appears in the second order solution $\delta^{[2]}_\lgr(\vq)$. Therefore we have $b_{K^2}= b_{K^2}^* D_*^{2}(\tau)$ plus terms proportional to $b_1^* -1$, in agreement with \refeq{bK2}.

Now consider a higher order non-factorizable operator $O_\lgr^{[n]}$ in the perturbative expansion of $O$. It is multiplied by $b_O(\tau) D^n(\tau)$ which is time-dependent. To make the entire contribution proportional to $O_\lgr^{[n]}$ time-independent and thus equal to its value at $\tau_*$, there needs to be an associated Eulerian operator $\hat O^{[n]}(\vx,\tau)=D^n(\tau)O_\lgr^{[n]}(\vq)+\O(\delta^{n+1})$ whose bias parameter satisfies
\be
\frac{\dc}{\dc\tau}\left[b_{\hat O^{[n]}}(\tau) D^n(\tau)+b_O(\tau) D^n(\tau)\right]=0.
\ee
This is solved by
\be
\label{eq:bObi}
b_{\hat O^{[n]}}(\tau)= b^*_{\hat O^{[n]}}D_*^n(\tau) +b_O(\tau)[D_*^{n-d_O}(\tau)-1]\,.
\ee
We see that even if initially $b_{\hat O^{[n]}}^* =0$, after a few Hubble times $b_{\hat O^{[n]}}(\tau)$ becomes of order $b_O(\tau)$. In practice there is a lot of overlap among various $n^{th}$ order operators. For instance, both $\delta_\lgr^{[3]}$ and $(K^2)_\lgr^{[3]}$ contain a piece proportional to $\Gamma^{[3]}_\lgr$. Again, superposition principle allows us to use the above method to calculate various contributions separately. For example:
\be
\label{eq:bnlK}
\frac16 b_{\Gamma}^E(\tau) \supset \frac{10}{21} b^E_{K^2}(\tau) [D_*(\tau)-1]\,,
\ee
where $10/21$ is the factor multiplying the term \eqref{eq:nl} in $(K^2)^{[3]}$ (note that $n$-th order biases are defined with a factor of $1/n!$). This agrees exactly with \refeq{bEnl}.
In order to understand the part proportional to $b_1^*-1$ in the latter equation, one has to take into account the indirect contribution from the second term on the r.h.s. of \refeq{bK2} {as well as the contribution from $\d$ at third order}.

\subsection{Formation history and age}

As argued above one can remain agnostic and use the most general basis \eqref{eq:listP}, or its Eulerian analog, with coefficients as free parameters to be determined empirically. However, it is natural to ask if there is any correlation between formation history or the age of galaxies and their bias parameters, given the relations \refeqs{b_1E}{bEnl}.

As mentioned earlier, the time derivatives must be accompanied by a time scale $T$. If the formation time is very fast perhaps it means $T\ll \cH_*^{-1}$, where $\cH_* = \cH(\tau_*)$. Naively, each time derivative should be suppressed by a power of $T\cH_*\ll 1$ at $\tau_*$. Hence, the operators such as $\Gamma$ which are defined using time derivatives would be expected to be initially suppressed, and to be generated mainly by the time-evolution (as in the second term on the r.h.s. of \refeq{bObi}). As such, their bias coefficients would be fixed in terms of other operators. However to define $\Phi$ we have already introduced a time-scale $\cH^{-1}$. To be consistent one should also replace that with $T$, resulting in a universal $(T\cH_*)^n$ suppression at order $n$ in perturbative expansion. Therefore, regardless of how fast the formation is, we do not expect time derivatives to be suppressed compared to other operators of the same order.

If we knew the formation history, that is, the values of $b_O^*$ at every $\tau_*$, we could obtain the final $\{b_O(\tau)\}$ by integrating the solution for instantaneous formation (\ref{eq:b_1E}-\ref{eq:bEnl}) over $\tau_*$. On the other hand, unless $T\gg \cH_*^{-1}$ or there are multiple far-separated events in the formation history of the objects of interest, this is unnecessary. At any order $n$ in perturbation theory time integration produces a finite number of non-zero bias coefficients for $\hat O^{[m]}$ [the Eulerian analog of $O_\lgr^{[m]}(\vq)$], with $m\leq n$. They can be expressed in terms of a finite number of time derivatives. Hence these objects can be approximated by instantaneously formed ones with small corrections to $b_{\hat O^{[m]}}^*$, $m\leq n$.

For very old galaxies, $D_*\ll 1$ and all Eulerian bias coefficients appear to be dominated by the piece proportional to $b_1^*-1$. In practice, $D_*$ is typically not too small.  Further, since at fixed Eulerian $b_1^E$ a small $D_*$ implies a very large $b_1^*$, the non-linear bias coefficients at the formation time may be correspondingly enhanced. Thus different contributions are generically comparable.\footnote{There is some evidence \cite{Baldauf,Sheth} for certain connections among bias coefficients of massive halos: $b_2\sim (b_1-1)^2$, $b_{K^2}\sim (b_1-1)$. This hierarchy between $\delta^2$ and $K^2$ does not seem to arise from our effective description. A possible explanation, proposed in \cite{Sheth}, is based on a generalization of the collapse model. In the original collapse model, $\delta_g \propto \exp(-\delta_c^2/\sigma^2(M))$, objects form when $\delta$ crosses a critical value $\delta_c$ and $\sigma^2(M)$ characterizes the variance of the initial density field. $\sigma(M)$ can become quite small for very massive objects.  One can generalize this model by replacing the collapse criterion $\delta=\delta_c$ with a {more general function of $\der_i\der_j\Phi$ with comparable coefficients $a_{\delta^2}\sim a_{K^2}$ for $\delta^2$ and $K^2$. What leads to a hierarchy in this approach is the rarity of halos at high masses due to the small variance $\sigma^2(M)\ll 1/a_{\delta^2}$ of the initial distribution}, since now $b_2 = \der^2\delta_g/\der\delta^2 \propto \delta_c^2/\sigma^4(M)+a_{\delta^2}/\sigma^2(M)$ while {$b_{K^2}\propto a_{K^2}/\sigma^2(M)$}.\label{b2K}}

\subsection{Velocity bias}
\label{sec:velbias}

{So far, we have assumed that tracers and matter comove on the same
fluid trajectories [\refeq{sdef}].  We now show why this is consistent at
the order in spatial derivatives we work in.}  
Apart from the universal gravitational force, tracers can in general experience different tidal forces $\vf_g$ compared to dark matter. The additional force  must again be a locally measurable quantity (in the absence of entropic perturbations).  The Euler equation for the tracer velocity field $\vv_g$ is then
\be
\partial_\tau\vv_g +\vv_g\cdot\nabla \vv_g = -\cH \vv_g -\nabla\phi+\vf_g\,.
\label{eq:Eulerg}
\ee
This is the same as the equation for $\vv$ in the EFT for dark matter \cite{Baumann} except that the back-reaction terms do not have to coincide, i.e. $\vf_g^i\neq \vf^i=\der_j\tau^{ij}$.  Subtracting the Euler equation from \refeq{Eulerg} and defining $\vv_{\rm rel} = \vv_g - \vv$, we arrive after some manipulation at the following system for conserved tracers' density contrast and velocity field:
\be
\begin{split}
\frac{\Dc}{\dc\tau}\vv_{\rm rel} + \cH \vv_{\rm rel} + \vv_{\rm rel}\cdot\nabla_x\vv_g = \vf_g - \vf\,,& \label{eq:vrel} \\
\frac{\Dc}{\dc\tau}(\delta_g-\delta) = -\theta(\delta_g-\delta)-\nabla\cdot [(1+\delta_g)\vv_{\rm rel}]\,.&
\end{split}
\ee
As a locally observable quantity, $\v{f}_g$ should be composed of the tidal tensor $\der_i\der_j\Phi$ and its derivatives. Expanding $\vf_g$ in powers of $\Phi$, isotropy implies that each term involving $n$ $\Phi$ fields must have at least $2n+1$ derivatives.  
It follows from \refeq{vrel} that $\vv_{\rm rel}$ can differ from zero only because of higher derivative force terms $\vf_g-\vf$ which would modify the bias coefficients for higher derivative operators, i.e. operators with at least two additional derivatives over the ones considered here.  Thus, our approach of setting $\vv_g = \vv$ throughout this paper is consistent.

\section{Connection with Lagrangian perturbation theory}
\label{sec:Lagr}

In this section we present a unified Lagrangian (LPT) picture by showing that all invariant quantities can be expressed using the distortion tensor $M^{ij}(\vx,\tau)=\der_q^i s^j(\vq,\tau)$ and its spatial and convective temporal derivatives. We then present a systematic, order by order construction of independent operators for a consistent bias expansion at any given order in perturbation theory.  

Showing the equivalence of the {LPT language and the (convective) SPT languague of the last section} is easy. The locally observable quantities such as
\be
\frac{\der v_i}{\der x_j} \qquad\mbox{and}\qquad \frac{\der^2\Phi}{\der x^i \der x^j}\,,
\ee
which we used so far, can be written in terms of $M$ and its time derivatives.  
We have
\be
v_i = \dot s_i\,, 
\ee
where here and in the following a dot stands for the convective derivative
with respect to $\tau$, i.e. $\dot{(\  )} = \Dc(\ )/\dc\tau$. [Recall that it reduces to ordinary derivative on functions of $(\vq,\tau)$.] This implies that $\der_i v_j$ is a measure of the time derivative of $M_{ij}$:
\be
{\partial v_i\over \partial x_j} = {\partial q_k\over \partial x_j} {\partial \dot s_i\over \partial q_k}.
\ee
We can compute 
\be
 {\partial q_k\over \partial x_j} = {\epsilon_{kmn} \epsilon_{jpl} \over 2 J} {\partial x_p\over \partial q_n}{\partial x_l\over \partial q_m}
\ee
with $J={\rm{det}}(\partial x/ \partial q) = {\rm{det}}(1+M)$, so that
\bea
{\partial v_i\over \partial x_j} &=& {\epsilon_{kmn} \epsilon_{jpl} \over 2 J} {\partial x_p\over \partial q_n}{\partial x_l\over \partial q_m} \dot M_{ik} \nonumber \\
&=& {\epsilon_{kmn} \epsilon_{jpl} \over 2 J} (\delta_{pn} + M_{pn})(\delta_{lm} + M_{lm}) \dot M_{ik}.
\eea
For the potential  on the other hand we can use
\be
\ddot s_i + {\cal H}  \dot s_i = - {\partial \phi \over \partial x_i},
\ee
so that 
\bea
{\partial^2 \phi \over \partial x_i \partial x_j} &=& -{\partial q_k\over \partial x_j} (\ddot M_{ik} + {\cal H}  \dot M_{ik}) \nonumber \\
&=& {\epsilon_{kmn} \epsilon_{jpl} \over 2 J} (\delta_{pn} + M_{pn})(\delta_{lm} + M_{lm}) (\ddot M_{ik} + {\cal H}  \dot M_{ik}).
\eea
Therefore, \emph{all terms allowed in the bias relations can be written using  $M$ and its time derivatives.} 

\subsection{Invariant density}

In order to gain insight into the general form of bias parametrization in terms of $M_{ij}$ it is useful to work out the explicit structure of Lagrangian operators $O_\lgr^{[n]}$ appearing in the perturbative expansion of some local operator $O(\vx,\tau)$ such as $\delta(\vx,\tau)$. The Lagrangian density components are defined in \refeq{delta}. The first observation is that one can easily get $\delta_{\lgr}^{[n]}(\q)$ using Lagrangian perturbation theory.  The density can be written as 
\be
1+\delta(x,\tau)= \left[{\rm det}(\delta_{ij}+M_{ij}) \right]^{-1}\Big|_{\x=\q+\s}.
\ee
The Lagrangian components $\delta_{\lgr}^{[n]}(\q)$ are nothing other than the perturbative expansion of the inverse determinant evaluated at $\q$. Let us write the determinant in terms of traces:
\be
{\rm det}(1+M) = 1 + {\rm  Tr}[M] -{1\over 2} ({\rm Tr}[M^2] - {\rm  Tr}[M]^2) + {1\over 6} (2 {\rm Tr}[M^3] - 3 {\rm Tr}[M^2] {\rm Tr}[M]+ {\rm Tr}[M]^3),
\ee 
where in perturbation theory we expand
\be
M(\vx,\tau)=D(\tau)M^{[1]}(\vq) + D^2(\tau) M^{[2]}(\vq) +D^3(\tau) M^{[3]}(\vq) + \cdots\,.
\nonumber
\ee
Also expanding the inverse determinant to third order, we get:
\bea
\delta &=& - D(\tau){\rm Tr}[M^{[1]}] + {1\over 2}D^2(\tau) \Big( {\rm Tr}[M^{[1]}]^2 + {\rm Tr}[M^{[1]}M^{[1]}] - 2 {\rm Tr}[M^{[2]}]\Big)  \nonumber  \\
& &+
D^3(\tau)\Big(-{{\rm Tr}[M^{[1]}]^3 \over 6} - { {\rm Tr}[M^{[1]}M^{[1]}] {\rm Tr}[M^{[1]}] \over 2} + {\rm Tr}[M^{[1]}] {\rm Tr}[M^{[2]}] - {{\rm Tr}[M^{[1]}M^{[1]}M^{[1]}] \over 3} \nonumber \\
& &+ {\rm Tr}[M^{[1]}M^{[2]}]- {\rm Tr}[M^{[3]}]\Big) + \cdots\,.
\eea
This yields
\bea
\delta_{\lgr}^{[1]}&=& - {\rm Tr}[M^{[1]}]   \nonumber \\
\delta_{\lgr}^{[2]}&=& {1\over 2} ( {\rm Tr}[M^{[1]}]^2 + {\rm Tr}[M^{[1]}M^{[1]}] - 2 {\rm Tr}[M^{[2]}])
 \nonumber \\
\delta_{\lgr}^{[3]}&=& (-{{\rm Tr}[M^{[1]}]^3 \over 6} - { {\rm Tr}[M^{[1]}M^{[1]}] {\rm Tr}[M^{[1]}] \over 2} + {\rm Tr}[M^{[1]}] {\rm Tr}[M^{[2]}] \nonumber \\ &&- {{\rm Tr}[M^{[1]}M^{[1]}M^{[1]}] \over 3} +{\rm Tr}[M^{[1]}M^{[2]}]- {\rm Tr}[M^{[3]}]).
\eea
To complete the dictionary at third order one needs to find $M^{[n]}$ for $n \leq 3$. Here we can use the recursive Lagrangian solution of \cite{Zheligovsky} (also quoted in \refeq{s_n} below). It is sufficient to know the gradient part of $M^{[n]}$, as the curl piece starts at third order and is odd under parity so it can only enter squared. The relevant solutions are:
\bea
{\rm Tr}[M^{[2]}]&=& {3\over 14} ({\rm Tr}[M^{[1]}M^{[1]}] - {\rm Tr}[M^{[1]}]^2)  \\
{\rm Tr}[M^{[3]}]&= &{1\over 63} (4 {{\rm Tr}[M^{[1]}]^3 } +3 { {\rm Tr}[M^{[1]}M^{[1]}] {\rm Tr}[M^{[1]}] } - 7 {{\rm Tr}[M^{[1]}M^{[1]}M^{[1]}] } +35 {\rm Tr}[M^{[1]}M^{[2]}])\nonumber
\eea
with
\be
M^{[2]}_{ij} = {\partial_i\partial_j \over \nabla^2} {\rm Tr}[M^{[2]}].
\ee
Thus at this order, the term  ${\rm Tr}[M^{[1]} M^{[2]}]$ is the sole new operator that maps to a time-derivative. In general, these are terms where $M^{[n]}$ with $n>1$ sits inside a trace multiplied by other matrices.

\subsection{Listing the bias parameters}
\label{sec:biasesM}

A complete list of operators can be obtained as in \refsec{flg}. All Lagrangian components of locally observable operators can be constructed by combining various $M^{[n]}(\vq)$ and taking traces. Thus the easiest way to count the number of free bias parameters up to a given order in perturbation theory is to count the number of different scalars that can be constructed combining $M^{[k]}$, $k=1, \cdots n$ where $k$ is the order in perturbation theory. A further simplification is that one never has to write ${\rm Tr} M^{[k]}$ because it can always be written in terms of scalars constructed using lower order objects. Up to including fourth order, the list reads\footnote{The number of operators grows very fast with order $n$. For instance the subset which only contains $M^{[1]}$ is identical to the number of partitions of $n$.}
\bea
{\rm 1^{st}} \ && \ {\rm Tr}[M^{[1]}]  \label{eq:listO} \\ 
{\rm 2^{nd}} \ && \ {\rm Tr}[(M^{[1]})^2],\  ({\rm Tr}[M^{[1]}])^2 \nonumber\\ 
{\rm 3^{rd}} \ && \ {\rm Tr}[(M^{[1]})^3 ],\ {\rm Tr}[(M^{[1]})^2 ]  {\rm Tr}[M^{[1]}],\ ({\rm Tr}[M^{[1]}])^3,\ {\rm Tr}[M^{[1]} M^{[2]}] \nonumber \\ 
{\rm 4^{th}} \ && \ {\rm Tr}[(M^{[1]})^4 ],\ {\rm Tr}[(M^{[1]})^3 ]  {\rm Tr}[M^{[1]}],\  {\rm Tr}[(M^{[1]})^2 ]  {\rm Tr}[(M^{[1]})^2],\ ({\rm Tr}[M^{[1]}])^4,\ {\rm Tr}[M^{[1]} M^{[3]}], \nonumber\\
&& \   {\rm Tr}[M^{[2]} M^{[2]}]. \nonumber
\eea
One can straightforwardly construct the analogous Eulerian basis using the full $M(\vx,\tau)$ and its time derivatives (by defining $\hat M^{[n]}$ operators analogous to $\hat\Pi^{[n]}$ of \refsec{flg}).

\section{Renormalization}

Another relevant question is how renormalization affects the above picture. The definition of non-linear operators that appear in the bias relation depends on a smoothing scale. Starting from one parametrization at a more refined level and smoothing over a range of short-wavelength modes will generate new operators of the smoothed field. This is due to the fact that the short-scale modes which are being averaged over evolve in the background of long-wavelength modes and hence their averages depend on various measurable quantities made of the long-wavelength modes. Moreover, since we are dealing with a pressureless fluid the short modes do not have a large frequency; they slowly evolve with characteristic time scale of $\cH^{-1}$. As a result the averages not only depend on observables of the long modes at the final time but also their value in the course of evolution along the trajectory of the short modes. This implies that, as in the case of the time evolution of bias parameters, the smoothing procedure will generate new operators which might have been absent originally. There are two questions to be asked at this point: (i) Are these new structures identical to those which arise from time evolution, or do they form a larger set? (ii) What is the characteristic size of the corrections? 

We leave a detailed analysis of renormalization of time evolving bias parameters for future work, and discuss only the qualitative features. Below in \refsec{rg} we will argue that the corrections to composite operators are within the same set of Lagrangian components $\{O_\lgr^{[n]}(\vq)\}$ of local operators. Hence by including time derivatives all counter-terms can be written in terms of local observables.

Unless there is a symmetry reason, loops will generate all of these operators. For instance, it was shown in \cite{Assassi} that starting from $\delta_g = b_1\delta+ b_2\delta^2$, in order to renormalize $\langle \delta_g \delta_1\delta_1\rangle$ at one-loop level one needs to add a new counter-term $K^2$, defined in \refsec{IF}. It is expected that the renormalization of $\langle \delta_g \delta_1\delta_1\delta_1 \rangle$ requires a non-zero bias parameter for $\Gamma$ operator.

As for the relative size of the corrections, it depends on how nonlinear the modes we are integrating out have become. At very large scales the corrections are expected to be small but at shorter scales they can become significant.

\begin{figure}[t]
\centering
\subfloat[][]{\includegraphics[scale=0.8]{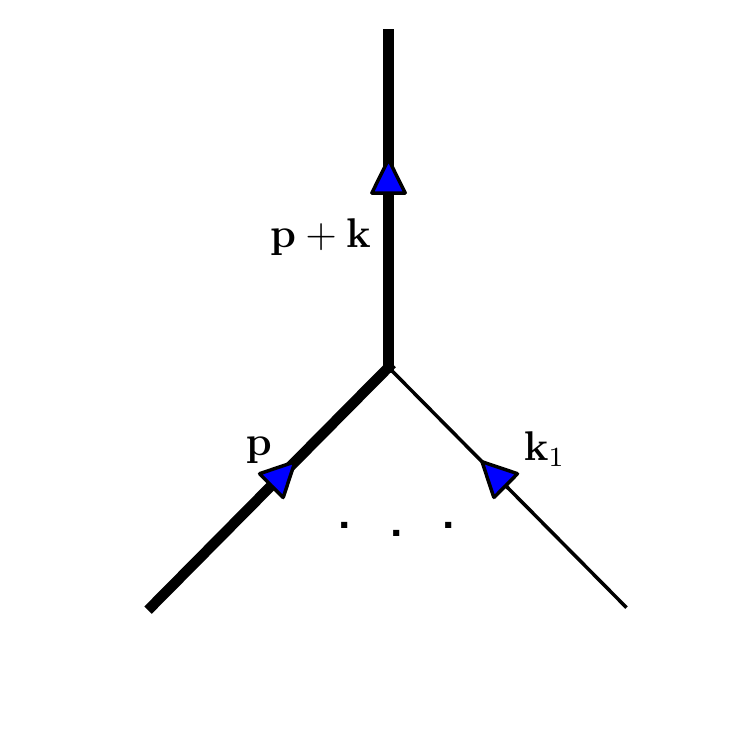}\label{hardout}}
\subfloat[][]{\includegraphics[scale=0.8]{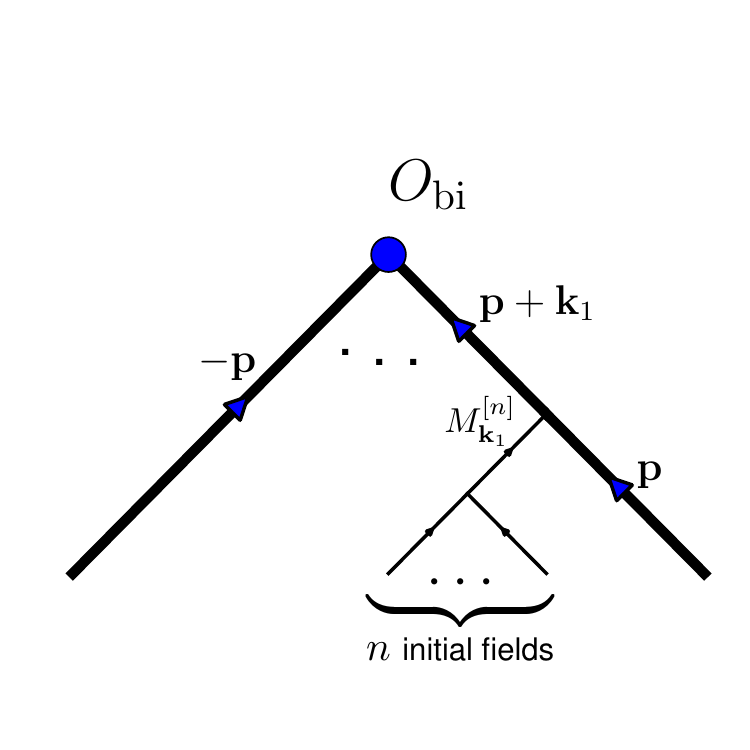}\label{hardin}}
\caption{\small{(a) A vertex with large incoming and outgoing momenta $p\sim \Lambda$. Dots represent omitted lines whose momenta add up to $\vk-\vk_1$. (b) A composite operator with two of the incoming momenta large and nearly opposite. All interactions along the large-momentum line yield local expressions in terms of the low-momentum incoming lines. }}
\end{figure}

\subsection{Renormalization and counter-terms}
\label{sec:rg}

As seen above the perturbation theory and bias parametrization can be fully formulated in terms of $M^{ij}=\der^i \s^j$. In this section we investigate the structure of counter-terms which must be added to the equations of motion of dark matter and to composite operators in order to remove the cutoff dependence of loop contributions to correlation functions.  

Here we are solving an initial value problem; at each vertex lower order solutions (incoming legs) combine to give a higher order solution (outgoing line). When calculating a correlation function of $\delta_g$ several incoming lines can end at a composite operator. Cutoff dependence arises when at least a pair of initial fields are contracted and their momenta become of the order of cutoff $\Lambda$. One can argue that the counter-terms are all local functions of $M^{ij}$ and its spatial or temporal derivatives:

\begin{enumerate}

\item The perturbation theory vertices express the trace and curl of $M^{[n]}$ as a local function of lower order solutions \cite{Zheligovsky}:
\be
\begin{split}
\left(\nabla\times \s^{[n]}\right)^i = &\frac{1}{2}\sum_{0<m<n}\frac{n-2m}{n}
\eps^{ijk} \left(M^{[m]} M^{[n-m]\,T}\right)_{jk}\,,\\
\nabla\cdot \s^{[n]} =& \frac{1}{2}\sum_{0<m<n}\frac{m^2+(n-m)^2+(n-3)/2}{n^2+(n-3)/2} 
[\Tr (M^{[m]} M^{[n-m]})-\Tr(M^{[m]})\Tr (M^{[n-m]})]\\
&-\frac{1}{6}\sum_{m_1+m_2+m_3=n}\frac{m_1^2+m_2^2+m_3^2+(n-3)/2}{n^2+(n-3)/2}
\vep^{ijk}\vep^{lmn}M^{[m_1]}_{il}M^{[m_2]}_{jm}M^{[m_3]}_{kn}\,.
\end{split}
\label{eq:s_n}
\ee
Non-localities appear when $M^{[n]}=\nabla \s^{(n)}$ is solved in terms of its symmetric and anti-symmetric parts:
\be
\s^{[n]}= \nabla^{-2}(-\nabla\times \nabla\times \s^{[n]}+\nabla \nabla\cdot \s^{[n]})\,.
\ee
Thus, $M^{[n]}_{ij}$ is always given by $\partial_i\partial_j/\nabla^2$ acting
on a product of $M^{[m]}$ with $m<n$.

\item  If the outgoing line has a large momentum (this can happen except for the final vertex or the external operator), the $\nabla^{-2}$ operator can be Taylor expanded in powers of external soft momenta and becomes local. For instance, in the diagram of Fig. \ref{hardout} the inverse Laplacian leads to a factor of $1/|\vp+\vk|^2$ which can be expanded in $k^2/p^2$. 

\item When two or more high-momentum internal lines end in a finite momentum composite operator in $\delta_g$ one needs to add a counter-term to renormalize the operator. This counter-term depends on the soft initial fields in the diagram. Because of the previous point it can be written as a product of various $M^{[n]}$, and their derivatives suppressed by factors of $1/\Lambda$ (Fig. \ref{hardin}). As before, these can be expressed in terms of $M$ its spatial, and convective time derivatives. The same argument also applies to interaction vertices of the dark matter field. 

\item Renormalization should be done order by order: the above argument works only if all nested diagrams with several $\O(\Lambda)$ momenta combining into soft momenta are already renormalized by the addition of appropriate counter-terms at lower orders.  

\end{enumerate}

To summarize, all Lagrangian counter-terms necessary in order to renormalize a local operator are those in the list \eqref{eq:listO} (and their higher spatial derivatives). Therefore, local functions of $M^{ij}$ as well its spatial and convective time derivatives form a complete basis for the renormalized bias expansion.

\section{Conclusions}
\label{sec:concl}

We argued that galaxy bias can be formulated locally at the final time in an Eulerian fashion in terms of local observables given by contractions of $\partial_i\partial_j\Phi$ and its convective time derivatives. Time derivatives are necessary to include starting from the third order. Equivalently one can use a Lagrangian description, e.g. in terms of $M^{ij}=\der_q^i s^j$. There is one bias parameter corresponding to each scalar quantity that can be constructed from product of various terms $M^{[n]}_{ij}$ appearing in the perturbative solution of $M_{ij}$. There is an analogous Eulerian operator for everyone of those. The complete list up to fourth order is given in \refeq{listP} and \refeq{listO}. Then, when including all bias parameters in this list at a given order, ``Eulerian'' and ``Lagrangian'' biasing are merely different equivalent formulations with the same physical content.
These claims were checked in an explicit third order calculation. Crucially, we also argued that the list of operators is closed under renormalization. The same rules can be used to construct a basis for the counter-terms in the EFT of the dark matter fluid.

A corollary of these considerations is that velocity bias can only enter
through higher spatial derivative terms.  At lowest order in spatial
derivatives, tracers and matter have to be comoving in the absence of
long-wavelength entropic perturbations.  {In other words, velocity biases of the type $\vv_g = b_v \vv + b_{v\d} \d\, \vv$ and
so on are prohibited by the equivalence principle in the absence of entropic
perturbations (which however can exist in standard cosmology for example in the form
of baryon-dark matter relative velocities \cite{tseliakhovich/hirata:2010}).}

Up to third order the proposed list of operators agrees with \cite{mcdonald/roy:2009,chan/scoccimarro/sheth:2012,Assassi}. However, beyond that one needs to include higher time derivatives and just working with local expressions made of gravitational and velocity potentials would be insufficient. On the other hand, our parametrization is more concise than that of \cite{Senatore:2014eva}.  Promoting all the bias coefficients to time integrals over history leads to redundancies in a perturbative approach.\footnote{The fact that the basis of operators in \cite{Senatore:2014eva} might be redundant was pointed out in that paper.}  Only a subset of operators involving convective time derivatives are required at any given order to build a complete basis. 

\section*{Acknowledgments}

We thank Tobias Baldauf, Daniel Baumann, Daniel Green, Enrico Pajer, and Leonardo Senatore for useful discussions. M.M. is supported by NSF Grants PHY-1314311 and PHY-0855425. M.Z. is supported in part by the NSF grants AST-0907969 and PHY-1213563.

\appendix

\section{Connection between velocity dispersion and tidal tensor}\label{app:dv}

Here we show that the velocity dispersion $\der_iv_j$ can be expressed perturbatively in terms of $\der_i\der_j\phi$ and its convective time derivatives. We start by taking a spatial derivative of the Euler equation:
\be
\partial_\tau \der_i v_j +(\vv\cdot\nabla) \der_iv_j +\der_i v_k \der_k v_j+ \cH \der_i v_j =-\der_i\der_j\phi\,.
\label{eq:Euler}
\ee
Next, we substitute the perturbative solutions for $\der_i\der_j\phi$ and $\der_iv_j$ in terms of Lagrangian operators $(\der_i\der_j\phi)_\lgr^{[n]}(\vq)$ and $(\der_iv_j)_\lgr^{[n]}(\vq)$. The convective time derivative on the l.h.s. reduces to an ordinary time derivative and we obtain an algebraic equation which can be solved perturbatively for $(\der_iv_j)_\lgr^{[n]}$ in terms of $(\der_i\der_j\phi)_\lgr^{[m]}$ with $m\leq n$. The latter can be solved in terms of zeroth to $n-1^{st}$ convective time-derivatives of $\der_i\der_j\phi(\vx,\tau)$ and therefore the same is true for $\der_iv_j(\vx,\tau)$ up to any order $n$. 

\section{Tracer with instantaneous formation at third order}
\label{app:convSPT}

The ``convective SPT'' system \refeqs{cont2}{cont1} can be written in compact form as
\ba
\frac{\Dc}{\dc\tau} \v{\Psi} =\:& - \v{\sigma}\cdot\v{\Psi} + \v{S} 
\quad\mbox{where}
\label{eq:Psigeom}\\
\v{\Psi}(\vx,\tau) = \left(\begin{array}{c}
\d_g(\vx,\tau) \\
\d(\vx,\tau) \\
\theta(\vx,\tau)
\end{array}
\right); \quad
\v{\sigma}(\tau) =\:& \left(\begin{array}{ccc}
0 & 0 & 1 \\
0 & 0 & 1 \\
0 & \frac32 \Om \cH^2 & \cH
\end{array}
\right); \quad
\v{S}(\vx,\tau) = \left(\begin{array}{c}
- \d_g\,\theta \\
- \d\,\theta \\
- (\partial_i \vv_j)^2
\end{array}
\right)\,.
\label{eq:Psigdef}
\ea
The fact that $\v{\sigma}$ is degenerate already shows that the three
equations are not really coupled, but rather the equation for $\d_g$ 
can be integrated separately.  However, \refeq{Psigeom}
allows for a convenient compact derivation of both $\d$ and $\d_g$.  
Note that since we integrate the system along the fluid flow, our 
expressions contain exclusively invariant operators.  For this
reason, we will drop the subscript ``bi'' in the majority
of this appendix for clarity.

\subsection{Source term}

Before we can integrate \refeq{Psigeom}, we have to carefully consider
the source term, in particular the third component $\v{S}_3 = (\partial_i \vv_j)^2$,
where the derivative is with respect to Eulerian coordinate $\vx$.  This
is nonlocally related to the degrees of freedom $\d,\,\theta$ themselves.  
For this, we use a trick (see also \refsec{Lagr}), 
namely the fact that $\vv$ is given by $\dot{\v{s}}$ (as in the main text,
dots stands for convective derivatives $\Dc/\dc\tau$).  
Further, we need to transform the
derivative from $\vx$ to the fluid flow or Lagrangian coordinate $\vq$ via
\be
\partial_x^i = \left[\v{1} + \v{M}\right]^{-1}{}^i_{\  j} \partial_q^j\,,
\ee
where $\v{M}^{ij} = \partial_q^i \v{s}^j$.  At second order, we then obtain
\be
\partial_x^i \vv^j = \left[ \partial_q^i + ( \partial_q^i \v{s}_k ) \partial_q^k \right] \dot{\v{s}}^j + \O(3)\,.
\ee
Here, the l.h.s is at $(\vx,\tau)$ while the r.h.s is at $(\vq,\tau)$
which are related through $\vx = \vq + \v{s}(\vq,\tau)$.  
We then obtain at first and second order
\ba
\partial_x^i \vv^j \Big|_{[1]} =\:& \partial_q^i \dot{\v{s}}_{[1]}^j \vs
\partial_x^i \vv^j \Big|_{[2]} =\:& \partial_q^i \dot{\v{s}}_{[2]}^j 
- (\partial_q^i \v{s}_{[1]}^k) \partial_{q, k} \v{s}_{[1]}^j\,.
\ea
As a consistency check of the second order result, we can easily verify
that the trace of the expression recovers $\theta^{[2]}_{\lgr}$, 
the invariant part of the velocity divergence at second order, i.e.
\be
\d_{ij} \partial_x^i \vv^j \Big|_{[2]} = - a\,\dot a \left[ \frac{13}{21} (\d^{[1]})^2
+ \frac47  (K_{ij}^{[1]})^2   \right]_{\tau_0}\,.  
\ee
With these expressions, we can construct source terms that are purely
in terms of Lagrangian derivatives.  Note that at each order, we only
need the displacement to one order less, since the source term is
quadratic.  
In the following, all spatial derivatives will be with respect to $\vq$.  
Pulling out $a$ factors for convenience, the third component of the
source term is at first and second order given by:
\ba
(\v{S}^{[1]})_3 =\:& - \dot a{}^2(\tau) \left[(K^{[1]})^2 + \frac13 (\d^{[1]})^2 \right]_{\tau_0} \vs
(\v{S}^{[2]})_3 =\:& 2 a\,\dot a{}^2(\tau) \left[
-\frac23 \d^{[1]} \sigma^{[2]} - 2 K^{[1]}_{ij} \Del^{ij} \sigma^{[2]} + \frac19 (\d^{[1]})^3 + \d^{[1]} (K^{[1]})^2 + (K^{[1]})^3 
\right]_{\tau_0}\,.
\label{eq:S1and2}
\ea
Here and in the following, we let $K^n$ stand for $\Tr[K_{ij}^n]$.  
$\sigma_{[2]} \equiv \vn_q\cdot \v{s}_{[2]}$ is given by
\ba
\sigma^{[2]} =\:& \frac12 \left[ -\frac27 (\d^{[1]})^2 + \frac37 (K^{[1]})^2 \right]
\,.
\label{eq:sigma2L}
\ea
These are sufficient to obtain second and third order results, respectively.

\subsection{Solution}

We define the matrix $\v{A}$ as the solution to the matrix ODE with boundary condition
\be
\partial_\tau \v{A}(\tau, \tau_i) + \sigma(\tau) \v{A}(\tau,\tau_i) = 0;\quad
\v{A}(\tau,\tau) = \v{1}\,.
\ee
For matter, we seek the solution to the subset $i=2,3$ of \refeq{Psigeom} 
with inital conditions at $\tau \to 0$ given by linear theory.  
This yields
\be
\left[\v{\Psi}\right]_{i=2,3}(\vx,\tau) = 
\left[\v{\Psi}^{[1]}\right]_{i=2,3}(\vx, \tau)
+ \int_0^\tau \dc\tau'\:\left[\v{A}(\tau,\tau')
\v{S}(\xfl(\tau'),\,\tau')\right]_{i=2,3} \,.
\label{eq:Psisol}
\ee
The linear solution for matter is given by
\be
\left[\v{\Psi}^{[1]}\right]_{i=2,3}(\vx, \tau) = \left(
\begin{array}{c}
a(\tau) \d_{[1]}(\xfl(0)) \\
-\dot a(\tau) \d_{[1]}(\xfl(0))
\end{array}\right)
= \left(
\begin{array}{c}
a(\tau) \\
-\dot a(\tau) 
\end{array}\right) \d_1(\vq)\,,
\ee
where $\d_{1}(\vq)$ is the linear density field as function of the Lagrangian
coordinate, extrapolated to the time $\tau_0$ where $a(\tau_0)=1$ following
standard convention.  As we will see below, \refeq{Psisol} leads to
the standard SPT result for matter for the \emph{invariant} terms,
while we have not expanded in the displacement yet.  Expanding the
displacements to the same perturbative order then yields exactly
the non-invariant (displacement) terms of SPT.

We now consider the conserved tracer at $\tau \geq \tau_*$.  The initial
condition for the integration of the full set of equations \refeq{Psigeom} 
is given by \refeq{dgIC},
\ba
\v{\Psi}(\vx_*, \tau_*) = \left(\begin{array}{c}
\d_g^* \\
\d^*\\
\theta^*
\end{array}\right)\,;\quad
\d_g^* =\:& \sum_{n=1}^3 \frac{b_n^*}{n!} [\d^*]^n 
+ \sum_{n=2}^3 \frac{b_{K^n}^*}{n!} \Tr \left[ (K_{ij}^*)^n \right]
+ \frac16 b_{\d K^2} \d^*\: \Tr \left[ (K_{ij}^*)^2 \right] \vs
& + \eps^*_0 + \eps^*_{\d} \d^* + \frac12 \eps^*_{\d^2} [\d^*]^2
+ \frac12 \eps^*_{K^2} \Tr \left[ (K_{ij}^*)^2 \right] 
\,,
\label{eq:PsigIC}
\ea
where a superscript $*$ indicates that a quantity is evaluated at
$\vx_* = \vx_{\rm fl}(\tau_*)$ and $\tau_*$.  
The quantities $\d^*,\,\theta^*$ are obtained by integrating \refeq{Psisol}
up to $\tau_*$.  
The particular solution to \refeq{Psigeom} with \refeq{PsigIC} is then given by
\be
\v{\Psi}(\vx,\tau) = 
\v{A}(\tau, \tau_*) 
\left(\begin{array}{c}
\d_g^* \\
\d^*\\
\theta^*
\end{array}\right)
+ \int_{\tau_*}^\tau \dc\tau'\:\v{A}(\tau,\tau')
\v{S}(\vx_{\rm fl}(\tau'),\,\tau') \,.
\label{eq:Psigsol}
\ee
The interpretation of \refeq{Psigsol} is clear:  the
density and velocity of the fluid and tracer at position $(\vx,\tau)$ is given by
an integral of the source term over the fluid trajectory.  

In the following, we assume an Einstein-de Sitter Universe ($\Om=1$), which yields
\be
\v{\sigma}(\tau) = \left(\begin{array}{ccc}
0 & 0 & 1 \\
0 & 0 & 1 \\
0 & \frac32 H_0^2 a^{-1} & H_0 a^{-1/2}
\end{array}
\right)
= \left(\begin{array}{ccc}
0 & 0 & 1 \\
0 & 0 & 1 \\
0 & 6 / \tau^2 & 2 / \tau
\end{array}
\right)\,.
\ee
The expression for $\v{A}$ is easy to derive since we know that the growing and decaying modes of the linear density field corresponding to the homogeneous solution
are $\propto \tau^2$ and $\tau^{-3}$, respectively.  
The linear solution is given by\footnote{Note that momentum conservation implies that the stochasticity in the matter density $\d$ is higher order in spatial derivatives, and we consequently neglect it.}
\be
\v{\Psi}^{[1]}(\vx, \tau) = a(\tau) \left(
\begin{array}{c}
b_1^E(\tau)\\
1 \\
- \cH(\tau) 
\end{array}\right)
\d_1(\vq)
+ \left(\begin{array}{c}
\eps_0^* \\
0\\
0
\end{array}\right)
\,,
\label{eq:Psig1}
\ee
where 
\be
b_1^E(\tau) \equiv (b_1^* -1) \frac{a(\tau_*)}{a(\tau)} + 1\,.
\label{eq:b1E}
\ee
We can then perform the usual
replacement of $a(\tau) \to D(\tau)$ to move from Einstein-de Sitter
to $\Lambda$CDM.  Thus, 
\be
\frac{a(\tau_*)}{a(\tau)} \to \frac{D(\tau_*)}{D(\tau)} \equiv D_*\,.
\ee

\subsection{Second and third order results}

Since the linear solution for $(\vx,\tau)$ where $\vx = \vx_{\rm fl}(\tau)$
is a function of $\vq = \vx_{\rm fl}(0)$, the second order source term at
$(\vx_{\rm fl}(\tau'),\tau')$ integrated over in \refeq{Psigsol} is only a 
function of $\vq$ and $\tau'$, so that the integral can be straightforwardly
evaluated.  This in fact holds order by order.  At second order, we have
\be
\v{S}^{[2]}(\vx_{\rm fl}(\tau'),\,\tau') =
\left(\begin{array}{c}
- \d_g^{[1]} \theta^{[1]} \\
- \d^{[1]} \theta^{[1]} \\
- \left[(\dot K^{[1]})^2 + \frac13 (\dot\d^{[1]})^2 \right]
\end{array}\right)_{\tau',\,\vq}
\,.
\ee
Performing the integral yields
\be
\v{\Psi}^{[2]}(\vx,\tau) 
= \left(
\begin{array}{c}
b_1^E(\tau) \d^{[2]} + b_2^E(\tau) (\d^{[1]})^2/2 + b_{K^2}^E(\tau) (K_{[1]}^{ij})^2/2 +  \eps^E_{\d}(\tau) \d^{[1]} 
\\
a^2(\tau) \left[\frac{17}{21} (\d^{[1]})^2 + \frac27 (K_{[1]}^{ij})^2 \right] \\
-a(\tau)\dot a(\tau) \left[ \frac{13}{21} (\d^{[1]})^2 + \frac47 (K_{[1]}^{ij})^2 \right]
\end{array}
\right)_{\vq}\,,
\label{eq:Psig2}
\ee
where we have defined second order Eulerian bias parameters
\ba
b_2^E(\tau) =\:& b_2^* D_*^2 - \frac8{21} (b_1^*-1) D_* (D_*-1) \vs
b_{K^2}^E(\tau) =\:& b_{K^2}^* D_*^2 + \frac47 (b_1^*-1) D_* (D_*-1)\,,
\label{eq:bE2}
\ea
and the Eulerian stochasticity
\ba
\eps^E_\d(\tau) = \eps^*_\d D_* - (D_*-1) \eps^*_0
\,.
\label{eq:epsE2}
\ea
Note that $\eps^E_\d$ has one power of $D_*$ less than the other
second order Eulerian quantities, since $\eps^*_i$ are defined at $\tau_*$.  

As mentioned above, $\v{\Psi}^{[2]}(\vx,\tau)$ is a local function of the
matter fields and stochastic variables evaluated at $\vq = \vx_{\rm fl}(0)$, and we obtain only invariant terms.  
Moving to Eulerian position $\vx$, 
the second order displacement term then simply adds to all three components
of $\v{\Psi}^{[2]}$ through
\be
\v{s}_{[1]}^i(\vx,\tau) \partial_i \v{\Psi}^{[1]}|_\vx\,.
\ee
At third order, we have
\be
\v{S}^{[2]}(\vx_{\rm fl}(\tau'_,\,\tau') =
\left(\begin{array}{c}
- (\d_g^{[1]} \theta^{[2]} + \d_g^{[2]} \theta^{[1]})_{\tau'} \\
- (\d^{[1]} \theta^{[2]} + \d^{[2]} \theta^{[1]})_{\tau'} \\
2 a\,\dot a{}^2(\tau) \left[
-\frac23 \d^{[1]} \sigma^{[2]} - 2 K^{[1]}_{ij} \Del^{ij} \sigma^{[2]} + \frac19 (\d^{[1]})^3 + \d^{[1]} (K^{[1]})^2 + (K^{[1]})^3 \right]_{\tau_0}
\end{array}\right)_{\vq}
\,.\nonumber
\ee
Inserting this into \refeq{Psigsol}, we correspondingly recover the invariant part of
the third order matter density and velocity dispersion (see \refapp{delta3}),
\ba
\d^{[3]} =\:& \frac{341}{567} \d^3 + \frac{11}{21} K^2 \d + \frac29 K^3 - \frac49 K^{ij} \Del_{ij} \sigma^{[2]} \vs
\theta^{[3]} =\:& \frac{\dot D(\tau)}{D(\tau)}\left[\frac{71}{189} \d^3 + \frac57 K^2 \d + \frac23 K^3 - \frac43 K^{ij} \Del_{ij} \sigma^{[2]}\right] \,,
\label{eq:delta3}
\ea
where all quantities on the r.h.s. are linear and evaluated at $\tau$.  
Notice that the prefactor of the $\d^3$ term in $\d^{[3]}$ is 
exactly the third-order coefficient of the perturbative expansion of 
spherical collapse in Einstein-de Sitter, as expected.  
For $\d_g$, we obtain at third order
\ba
\d_{g}^{[3]}(\vx,\tau) =\:& b_1^E \d^{[3]} + b_2^E \d^{[1]} \d^{[2]} + b_{K^2}^E K^{[1]} K^{[2]} + \frac16 \left[b_3^E (\d^{[1]})^3 + b_{K^3}^E (K^{[1]})^3 + b_{\d K^2} \d^{[1]} (K^{[1]})^2 \right] \vs
& + \frac16 b_{\rm nloc}^E K^{[1]}_{ij} \Del^{ij} \left[(\d^{[1]})^2 
-\frac32 (K^{[1]})^2 \right] \vs
& + \eps_\d^E \d^{[2]} + \frac12 \eps_{\d^2}^E (\d^{[1]})^2 
+ \frac12 \eps_{K^2}^E (K^{[1]})^2 
\,,
\label{eq:dg3f}
\ea
where the third order final-time bias parameters are given by
\ba
b_3^E(\tau) =\:& b_3^* D_*^3 + \left[
(b_1^*-1) \frac4{1323} (199 - 35 D_*) 
+ \frac{13}7 b_2^* D_* 
\right]
D_* (D_*-1)
\vs 
b_{\d K^2}^E(\tau) =\:& b_{\d K^2}^* D_*^3 + \left[
- (b_1^*-1) \frac2{49} (33+7 D_*) 
+ \left(\frac{12}7 b_2^* + 3 b_{K^2}^*\right) D_*
\right]
D_* (D_*-1)
\vs
b_{K^3}^E(\tau) =\:& b_{K^3}^* D_*^3 + \left[
(b_1^*-1) \frac4{21} (-11 + 7 D_*) 
+ 6 b_{K^2}^* D_* 
\right]
D_* (D_*-1)
\vs
b_{\rm nloc}^E(\tau) =\:& \left[
(b_1^*-1) \left(-\frac8{21}\right) \left(\frac{23}7 - D_*\right) 
+\frac{20}7 b_{K^2}^* D_* 
\right]
D_* (D_*-1)
\,.
\label{eq:biasEthirdorder}
\ea
We can see that, as expected, all corrections to the bias parameters at $\tau_*$
disappear for $D_* \to 1$ and $D_* \to 0$ (unless $b_X^*$ diverge in the latter
limit, which is usually assumed for Lagrangian biasing).  
$b_{\rm nloc}^E$ multiplies third order terms which
cannot be expressed locally in terms of the linear density and tidal field.  
They are exactly of the structure discussed in \refsec{basics}.  
Moreover, using that 
\be
\Del_{ij} \sigma^{[2]}
= - \frac17 \Del_{ij} \left[ \d^2 - \frac32 K^2\right]\,,
\ee
we see that, to the order we work in, we can equivalently write \refeq{dg3f} as
\ba
\d_{g}^{[3]}(\vx,\tau) =\:& b_1^E \d^{[3]} + b_2^E \d^{[1]} \d^{[2]} + b_{K^2}^E K^{[1]} K^{[2]} + \frac16 \left[b_3^E (\d^{[1]})^3 + b_{K^3}^E (K^{[1]})^3 + b^E_{\d K^2} \d^{[1]} (K^{[1]})^2 \right] \vs
& + \frac16 \left(- 7 b_{\rm nloc}^E \right)  K_{[1]}^{ij} M^{[2]}_{ij}\,,
\ea
where
\be
M^{[2]}_{ij} = \frac{\partial_q^i \partial_q^j}{\nabla_q^2} \sigma^{[2]}
\ee
is the second order Lagrangian distortion tensor. 
This can be equivalently written in terms of $M'_{ij} - M_{ij}$ (see \refsec{biasesM}).  
The Eulerian stochasticity terms at third order become
\ba
\eps_{\d^2}^E(\tau) =\:& \eps^*_{\d^2} D_*^2 - \frac8{21} (\eps_\d^* - \eps_0^*)
D_* (D_*-1)
\vs
\eps_{K^2}^E(\tau) =\:& \eps^*_{K^2} D_*^2 + \frac47 (\eps_\d^* - \eps_0^*)
D_* (D_*-1)\,.
\label{eq:epsEthirdorder}
\ea
Finally, in order to obtain the density at a fixed order in standard Eulerian perturbation
theory, we need to expand the argument of the various orders of 
$\v{\Psi}$.  Let us define $\v{\Psi}^E$ through
\be
\v{\Psi}^E(\vx) = \v{\Psi}(\vx - \v{s}[\vq,\tau])\,.
\ee
$\v{\Psi}$ on the r.h.s. contains all invariant terms, which we have
worked out here.  We now perform a Taylor expansion in $\v{s}$
as well as a perturbative expansion of $\v{s}$, noting that $\v{s}$ is itself
a function of the Lagrangian rather than Eulerian position.  We obtain
\ba
\v{\Psi}^E 
=\:& \v{\Psi}^{[3]} - \v{s}_{[1]}^i \partial_i \v{\Psi}^{[2]}
- \left[\v{s}_{[2]}^i - \v{s}_{[1]}^j (\partial_j \v{s}_{[1]}^i) \right] \partial_i \v{\Psi}^{[1]} 
+ \frac12 \v{s}_{[1]}^i \v{s}_{[1]}^j \partial_i \partial_j \v{\Psi}^{[1]} \,,
\label{eq:Psigtotal}
\ea
where on the r.h.s. all quantities are evaluated at $\vx,\,\tau$, and
$\v{\Psi}^{[1]}$, $\v{\Psi}^{[2]}$, $\v{\Psi^{[3]}}$ are given in \refeq{Psig1}, \refeq{Psig2}, and \refeqs{delta3}{dg3f}, respectively.  

An alternative derivation of these results can be obtained as follows (see Sec.~III of Ref.~\cite{chan/scoccimarro/sheth:2012}).  For
conserved tracers, the continuity equation \refeq{cont1} and absence of velocity bias at lowest order
in derivatives (\refsec{velbias}) imply
\be
-\theta = \frac1{1+\d_g} \frac{\Dc \d_g}{\dc\tau} = \frac1{1+\d} \frac{\Dc \d}{\dc\tau} \,.
\ee
This can be immediately integrated to yield
\be
\ln [1 + \d_g(\xfl(\tau),\tau)] = \ln [1 + \d(\xfl(\tau),\tau)] + F[\xfl(\tau_*)]\,,
\label{eq:dgcons}
\ee
where $F$ is a function of space only, which we have chosen to be defined
on the formation time slice $\tau_*$.  Evaluating \refeq{dgcons} at $\tau_*$, we obtain
\be
F(\vx_*) = \ln \left[\frac{1 + \d_g(\vx_*, \tau_*)}{1+\d(\vx_*,\tau_*)}\right]\,.
\label{eq:Fx}
\ee
One can now insert the bias expansion \refeq{PsigIC} and expand \refeq{dgcons} to the desired order.

\section{Matter density perturbation at third order}
\label{app:delta3}

For reference, we now show how to obtain an expression for the third order 
density field in real space, with invariant and non-invariant terms
explicitly separated, from standard perturbation theory expressions.  
We start from the symmetrized third-order perturbation
theory kernel \cite{Bernardeau/etal:2002},
\ba
F_3(\vk_1,\vk_2,\vk_3) =\:& 
\frac2{54} k_{123}^2 \left[\frac{\vk_1\cdot\vk_{23}}{k_1^2 k_{23}^2} G_2(\vk_2,\vk_3) + 2\:\rm perm.\right] \vs
& + \frac7{54} \vk_{123} \cdot\left[\frac{\vk_{12}}{k_{12}^2} G_2(\vk_1,\vk_2) + 2\:\rm perm.\right] \vs
& + \frac7{54} \vk_{123} \cdot\left[\frac{\vk_{1}}{k_{1}^2} F_2(\vk_2,\vk_3) + 2\:\rm perm.\right] \,,
\label{eq:F3}
\ea
where $F_2$ and $G_2$ are given by
\ba
F_2(\vk_1,\vk_2) =\:& \frac57 + \frac27 \frac{(\vk_1\cdot\vk_2)^2}{k_1^2 k_2^2}
+ \frac12 \vk_1\cdot\vk_2 \left(k_1^{-2} + k_2^{-2}\right)
\vs
G_2(\vk_1,\vk_2) =\:& \frac37 + \frac47 \frac{(\vk_1\cdot\vk_2)^2}{k_1^2 k_2^2}
+ \frac12 \vk_1\cdot\vk_2 \left(k_1^{-2} + k_2^{-2}\right)\,.
\ea
Transforming the Fourier integral over \refeq{F3} into real space, we
obtain after some algebra
\ba
\d^{[3]} =\:& \frac{341}{567} \d^3 + \frac{71}{189} \d\,K^2 
- \frac{145}{189} \d\,\v{s}^i\partial_i \d
- \frac{11}{63} \v{s}^i\partial_i K^2
+ \frac12 \v{s}^i (\partial_i \v{s}^j) \partial_j \d
+ \frac12 \v{s}^i\v{s}^j \partial_i \partial_j \d \vs
& +\frac12 \frac{\partial^i}{\nabla^2} \left[\frac{13}{21} \d^2 + \frac47 [K_{jk}]^2 - \v{s}^j \partial_j \d \right] \partial_i \d 
+ \frac29 K_{ij} \Del^{ij} \left[\frac{13}{21} \d^2 + \frac47 [K_{ij}]^2 - \v{s}^j \partial_j \d \right]\,.
\label{eq:d3var1}
\ea
Here and in the following,
all terms without further sub/superscripts denote linear order quantities.  

We are interested in separating terms that involve the displacement
(non-invariant) and those that do not (invariant) in 
\refeq{d3var1}.  For this, we need to manipulate the terms which
contain a displacement inside the inverse Laplacian.  
Using a suitable reordering of $k$ factors in Fourier space, one can
easily prove the following relations:
\ba
\frac{\partial^i \partial^j}{\nabla^2} \left[ \v{s}^j \partial_j \d \right ]
=\:& - \frac12 \partial^i \partial^j [ \v{s}^2 ] + \frac{\partial^i \partial^j}{\nabla^2} \left[ K^2 + \frac13 \d^2 \right]
\vs
\frac{\partial^i}{\nabla^2} \left[ \v{s}^j \partial_j \d \right ]
=\:& -\frac12 \partial^i \left[\v{s}^2 \right] + \frac{\partial^i}{\nabla^2} \left[ K^2 + \frac13 \d^2 \right] 
= - \v{s}^k (\partial_k \v{s}^i)  + \frac{\partial^i}{\nabla^2} \left[ K^2 + \frac13 \d^2 \right]
\,,
\label{eq:nonloctransf}
\ea
where we have used $\v{s}^i = -(\partial^i/\nabla^2) \d$ to switch derivatives.  
This yields
\ba
K^{ij} \Del_{ij} \left[ \v{s}^j \partial_j \d \right ]
=\:& - K^3 - \frac23 \d K^2 + \frac12 \v{s}^k \partial_k K^2 + K^{ij} \Del_{ij} \left[K^2 + \frac13 \d^2 \right] \vs
-\frac{\partial^i}{\nabla^2} \left[\frac{13}{21} \d^2 + \frac47 [K_{jk}]^2 - \v{s}^j \partial_j \d \right] =\:& -\frac{\partial^i}{\nabla^2} \left[\frac27 \d^2 - \frac37 K^2\right] - \v{s}^k (\partial_k \v{s}^i)\,.
\label{eq:K2nonloctransf}
\ea
The second equation is the relation between second-order displacements 
at fixed Eulerian position (left-hand side), given by the time integral
over $-(\partial^i/\nabla^2) \theta^{[2]}$, and fixed Lagrangian position
(first term on the right-hand side), precisely $2\v{s}_{[2]}^i$.  
Inserting these relations into \refeq{d3var1}, we obtain
\ba
\d^{[3]} =\:& \d^{[3]}_{\lgr} +
\frac12 \frac{\partial^i}{\nabla^2} \left[\frac27 \d^2 - \frac37 K^2\right] \partial_i \d 
+ \v{s}^k (\partial_k \v{s}^i) \partial_i \d 
+ \frac12 \v{s}^i \v{s}^k \partial_i  \partial_k \d 
- \v{s}^i \partial_i\left[ \frac{17}{21} \d^2 + \frac27 K^2 \right] 
\label{eq:d3var2} \\
\d^{[3]}_{\lgr} =\:&
\frac{341}{567} \d^3 + \frac{11}{21} \d K^2 + \frac 29 K^3 
+ \frac29 K^{ij} \Del_{ij} \left[ \frac27 \d^2 - \frac37 K^2  \right]\,,
\nonumber
\ea
where we have collected the invariant terms in $\d^{[3]}_{\lgr}$.  
The remaining
terms in \refeq{d3var2} are the displacement or non-invariant terms.  
Using the notation defined above [\refeq{sigma2L}], 
we can write this as
\ba
\d^{[3]} =\:& -\v{s}_{[2]}^i \partial_i \d 
+ \v{s}^k (\partial_k \v{s}^i) \partial_i \d 
+ \frac12 \v{s}^i \v{s}^k \partial_i  \partial_k \d 
- \v{s}^i \partial_i \d^{[2]}_{\lgr} \vs
& + \frac{341}{567} \d^3 + \frac{11}{21} \d K^2 + \frac 29 K^3 
- \frac49 K^{ij} \Del_{ij} \sigma_{[2]}\,,
\label{eq:d3var3}
\ea
which agrees with \refeq{Psigtotal} and \refeq{delta3} from the convective
SPT approach.

\bibliography{msz}

\end{document}